\documentclass[twocolumn, preprint]{aastex62}
\graphicspath{{./}{figures/}}
\usepackage{subfiles}
\usepackage{lipsum}
\usepackage{natbib}
\received{April 1, 2020}
\revised{\today}
\accepted{``Frabjous day Callooh! Callay!'' }

\submitjournal{ApJ}
\shorttitle{Analysis of HAWC~J2019+368}
\shortauthors{Brisbois, Joshi et al.}

\newcommand{\gam}{$\gamma$}

\begin{document}

\title{Spectrum and Morphology of the Very-High-Energy Source HAWC~J2019+368} 
\def \UMD {Department of Physics, University of Maryland, College Park, MD, USA }
\def \ECAP {Erlangen Centre for Astroparticle Physics, Friedrich-Alexander-Universit\"at Erlangen-N\"urnberg, Erlangen, Germany}

\correspondingauthor{Chad Brisbois}
\email{chadb@umd.edu}
\affiliation{\UMD}

\correspondingauthor{Vikas Joshi}
\email{vikas.joshi@fau.de}
\affiliation{\ECAP}

%
%
% Keep the rest of the Authors down here
%
%

\def \LANL {Physics Division, Los Alamos National Laboratory, Los Alamos, NM, USA}

\def \IFUNAM {Instituto de F\'{i}sica, Universidad Nacional Aut\'onoma de M\'exico, Ciudad de Mexico, Mexico}

\def \UNACH {Universidad Aut\'onoma de Chiapas, Tuxtla Guti\'errez, Chiapas, M\'exico}

\def \UMSNH {Universidad Michoacana de San Nicol\'as de Hidalgo, Morelia, Mexico}

\def \PSU {Department of Physics, Pennsylvania State University, University Park, PA, USA}

\def \INAOE {Instituto Nacional de Astrof\'{i}sica, \'Optica y Electr\'onica, Puebla, Mexico}

\def \IFJPAN {Institute of Nuclear Physics Polish Academy of Sciences, PL-31342 IFJ-PAN, Krakow, Poland}

\def \FCFMBUAP {Facultad de Ciencias F\'{i}sico Matem\'aticas, Benem\'erita Universidad Aut\'onoma de Puebla, Puebla, Mexico}

\def \UdG {Departamento de F\'{i}sica, Centro Universitario de Ciencias Exactase Ingenierias, Universidad de Guadalajara, Guadalajara, Mexico}

\def \UWMadison {Department of Physics, University of Wisconsin-Madison, Madison, WI, USA}

\def \IAUNAM {Instituto de Astronom\'{i}a, Universidad Nacional Aut\'onoma de M\'exico, Ciudad de Mexico, Mexico}

\def \MPIK {Max-Planck Institute for Nuclear Physics, 69117 Heidelberg, Germany}

\def \MTU {Department of Physics, Michigan Technological University, Houghton, MI, USA}

\def \MSU {Department of Physics and Astronomy, Michigan State University, East Lansing, MI, USA}

\def \UPP {Universidad Politecnica de Pachuca, Pachuca, Hgo, Mexico}

\def \INFNPadova {INFN and Universita di Padova, via Marzolo 8, I-35131,Padova,Italy}

\def \CICIPN {Centro de Investigaci\'on en Computaci\'on, Instituto Polit\'ecnico Nacional, M\'exico City, M\'exico.}

\def \UNM {Dept of Physics and Astronomy, University of New Mexico, Albuquerque, NM, USA}

\def \ICNUNAM {Instituto de Ciencias Nucleares, Universidad Nacional Aut\'onoma de Mexico, Ciudad de Mexico, Mexico}

\def \CUBT {Department of Physics, Faculty of Science, Chulalongkorn University, 254 Phayathai Road, Pathumwan, Bangkok 10330, Thailand}

\def \NARIT {National Astronomical Research Institute of Thailand (Public Organization), Don Kaeo, MaeRim, Chiang Mai 50180, Thailand}

\def \IGeof {Instituto de Geof\'{i}sica, Universidad Nacional Aut\'onoma de M\'exico, Ciudad de Mexico, Mexico}

\def \Stanford {Department of Physics, Stanford University: Stanford, CA 94305–4060, USA}

\def \UOS {University of Seoul, Seoul, Rep. of Korea}

\def \UofUtah {Department of Physics and Astronomy, University of Utah, Salt Lake City, UT, USA}

\def \UAEH {Universidad Aut\'onoma del Estado de Hidalgo, Pachuca, Mexico} 

\def \CALU {Department of Chemistry and Physics, California University of Pennsylvania, California, Pennsylvania, USA}

\def \STJU {Tsung-Dao Lee Institute \& School of Physics and Astronomy, Shanghai Jiao Tong University, Shanghai, China} 

\author[0000-0003-0197-5646]{A.~Albert}
\affiliation{\LANL}

\author[0000-0001-8749-1647]{R.~Alfaro}
\affiliation{\IFUNAM}

\author{C.~Alvarez}
\affiliation{\UNACH}

\author{J.C.~Arteaga-Vel\'azquez}
\affiliation{\UMSNH}

\author[0000-0002-3032-663X]{K.P.~Arunbabu}
\affiliation{\IGeof}

\author[0000-0002-4020-4142]{D.~Avila Rojas}
\affiliation{\IFUNAM}

\author[0000-0002-2084-5049]{H.A.~Ayala Solares}
\affiliation{\PSU}

\author[0000-0003-0477-1614]{V.~Baghmanyan}
\affiliation{\IFJPAN}

\author[0000-0003-3207-105X]{E.~Belmont-Moreno}
\affiliation{\IFUNAM}

\author[0000-0002-5493-6344]{C.~Brisbois}
\affiliation{\UMD}

\author[0000-0002-4042-3855]{K.S.~Caballero-Mora}
\affiliation{\UNACH}

\author[0000-0003-2158-2292]{T.~Capistr\'an}
\affiliation{\IAUNAM}

\author[0000-0002-8553-3302]{A.~Carrami\~nana}
\affiliation{\INAOE}

\author[0000-0002-6144-9122]{S.~Casanova}
\affiliation{\IFJPAN}

\author[0000-0002-1132-871X]{J.~Cotzomi}
\affiliation{\FCFMBUAP}

\author[0000-0002-7747-754X]{S.~Couti\~no de Le\'on}
\affiliation{\INAOE}

\author[0000-0001-9643-4134]{E.~De la Fuente}
\affiliation{\UdG}

\author{R.~Diaz Hernandez}
\affiliation{\INAOE}

\author[0000-0001-8451-7450]{B.L.~Dingus}
\affiliation{\LANL}

\author[0000-0002-2987-9691]{M.A.~DuVernois}
\affiliation{\UWMadison}

\author[0000-0003-2169-0306]{M.~Durocher}
\affiliation{\LANL}

\author[0000-0001-5737-1820]{K.~Engel}
\affiliation{\UMD}

\author[0000-0001-7074-1726]{C.~Espinoza}
\affiliation{\IFUNAM}

\author[0000-0002-0173-6453]{N.~Fraija}
\affiliation{\IFUNAM}

\author{D.~Garcia}
\affiliation{\IFUNAM}

\author[0000-0002-4188-5584]{J.A.~Garc\'ia-Gonz\'alez}
\affiliation{\IFUNAM}

\author{G.~Giacinti}
\affiliation{\MPIK}

\author[0000-0002-5209-5641]{M.M.~Gonz\'alez}
\affiliation{\IAUNAM}

\author[0000-0002-9790-1299]{J.A.~Goodman}
\affiliation{\UMD}

\author[0000-0001-9844-2648]{J.P.~Harding}
\affiliation{\LANL}

\author[0000-0002-1031-7760]{J.~Hinton}
\affiliation{\MPIK}

\author{B.~Hona}
\affiliation{\MTU}

\author[0000-0002-3808-4639]{D.~Huang}
\affiliation{\MTU}

\author[0000-0002-5527-7141]{F.~Hueyotl-Zahuantitla}
\affiliation{\UNACH}

\author{P.~Huentemeyer}
\affiliation{\MTU}

\author[0000-0002-6738-9351]{A.~Jardin-Blicq}
\affiliation{\MPIK}
\affiliation{\CUBT}
\affiliation{\NARIT}

\author[0000-0003-4467-3621]{V.~Joshi}
\affiliation{\ECAP}
\affiliation{\MPIK}

\author[0000-0002-2467-5673]{W.H.~Lee}
\affiliation{\IAUNAM}

\author[0000-0001-5516-4975]{H.~Le\'on Vargas}
\affiliation{\IFUNAM}

\author[0000-0003-2696-947X]{J.T.~Linnemann}
\affiliation{\MSU}

\author[0000-0001-8825-3624]{A.L.~Longinotti}
\affiliation{\INAOE}
\affiliation{\IAUNAM}

\author[0000-0003-2810-4867]{G.~Luis-Raya}
\affiliation{\UPP}

\author[0000-0003-3751-5617]{J.~Lundeen}
\affiliation{\MSU}

\author[0000-0002-3882-9477]{R.~L\'opez-Coto}
\affiliation{\INFNPadova}

\author[0000-0001-8088-400X]{K.~Malone}
\affiliation{\LANL}

\author[0000-0001-9052-856X]{O.~Martinez}
\affiliation{\FCFMBUAP}

\author[0000-0002-2824-3544]{J.~Mart\'inez-Castro}
\affiliation{\CICIPN}

\author[0000-0002-2610-863X]{J.A.~Matthews}
\affiliation{\UNM}

\author[0000-0002-8390-9011]{P.~Miranda-Romagnoli}
\affiliation{\UAEH}

\author{J.A.~Morales-Soto}
\affiliation{\UMSNH}

\author[0000-0002-1114-2640]{E.~Moreno}
\affiliation{\FCFMBUAP}

\author[0000-0002-7675-4656]{M.~Mostaf\'a}
\affiliation{\PSU}

\author[0000-0003-0587-4324]{A.~Nayerhoda}
\affiliation{\IFJPAN}

\author[0000-0003-1059-8731]{L.~Nellen}
\affiliation{\ICNUNAM}

\author[0000-0001-9428-7572]{M.~Newbold}
\affiliation{\UofUtah}

\author[0000-0002-0479-2311]{M.U.~Nisa}
\affiliation{\MSU}

\author[0000-0001-7099-108X]{R.~Noriega-Papaqui}
\affiliation{\UAEH}

\author[0000-0002-9105-0518]{L.~Olivera-Nieto}
\affiliation{\MPIK}

\author[0000-0002-5448-7577]{N.~Omodei}
\affiliation{\Stanford}

\author{A.~Peisker}
\affiliation{\MSU}

\author[0000-0002-8774-8147]{Y.~P\'erez~Araujo}
\affiliation{\IAUNAM}

\author[0000-0001-5998-4938]{E.G.~P\'erez-P\'erez}
\affiliation{\UPP}

\author[0000-0002-6524-9769]{C.D.~Rho}
\affiliation{\UOS}

\author[0000-0003-1327-0838]{D.~Rosa-Gonz\'alez}
\affiliation{\INAOE}

\author[0000-0001-6939-7825]{E.~Ruiz-Velasco}
\affiliation{\MPIK}

\author{H.~Salazar}
\affiliation{\FCFMBUAP}

\author[0000-0002-8610-8703]{F.~Salesa~Greus}
\affiliation{\IFJPAN}
\affiliation{Instituto de F\'isica Corpuscular, CSIC, Universitat de Val\`encia, E-46980, Paterna, Valencia, Spain}

\author[0000-0001-6079-2722]{A.~Sandoval}
\affiliation{\IFUNAM}

\author[0000-0001-8644-4734]{M.~Schneider}
\affiliation{\UMD}

\author[0000-0002-8999-9249]{H.~Schoorlemmer}
\affiliation{\MPIK}

\author{J.~Serna-Franco}
\affiliation{\IFUNAM}

\author[0000-0002-1012-0431]{A.J.~Smith}
\affiliation{\UMD}

\author[0000-0002-1492-0380]{R.W.~Springer}
\affiliation{\UofUtah}

\author[0000-0002-8516-6469]{P.~Surajbali}
\affiliation{\MPIK}

\author[0000-0001-9725-1479]{K.~Tollefson}
\affiliation{\MSU}

\author[0000-0002-1689-3945]{I.~Torres}
\affiliation{\INAOE}

\author{R.~Turner}
\affiliation{\MTU}

\author[0000-0002-2748-2527]{F.~Ure\~na-Mena}
\affiliation{\INAOE}

\author{T.~Weisgarber}
\affiliation{\CALU}

\author{E.~Willox}
\affiliation{\UMD}

\author[0000-0003-0513-3841]{H.~Zhou}
\affiliation{\STJU}

\author[0000-0002-8528-9573]{C.~de Le\'on}
\affiliation{\UMSNH}

%\author[0000-0002-0786-7307]{Greg J. Schwarz}
%\affil{American Astronomical Society \\
%2000 Florida Ave., NW, Suite 300 \\
%Washington, DC 20009-1231, USA}

%\author{August Muench}
%\affiliation{American Astronomical Society \\
%2000 Florida Ave., NW, Suite 300 \\
%Washington, DC 20009-1231, USA}
\collaboration{HAWC Collaboration}

\begin{abstract}
The MGRO~J2019+37 region is one of the brightest sources in the sky at TeV energies. It was detected in the 2 year HAWC catalog as 2HWC J2019+367 and here we present a detailed study of this region using data from HAWC. This analysis resolves the region into two sources: HAWC~J2019+368 and HAWC~J2016+371. We associate HAWC~J2016+371 with the evolved supernova remnant CTB 87, although its low significance in this analysis prevents a detailed study at this time. An investigation of the morphology (including possible energy dependent morphology) and spectrum for HAWC~J2019+368 is the focus of this work. We associate HAWC~J2019+368 with PSR~J2021+3651 and its X-ray pulsar wind nebula, the Dragonfly nebula. Modeling the spectrum measured by HAWC and \textit{Suzaku} reveals a $\sim7$ kyr pulsar and nebula system producing the observed emission at X-ray and \gam-ray energies. 
\end{abstract}
\keywords{gamma rays: general -- pulsars: general -- PWN: general -- individual objects}
\section{Introduction}

The brightest source in the Cygnus region observed by the Milagro observatory was MGRO~J2019+37 \citep{Abdo_2007}. Milagro measured its flux to be about 80\% of the Crab Nebula flux at 20 TeV and showed an extent of $\sim\!\!0.75\degr$ \citep{Abdo_2012}. The MGRO~J2019+37 region was later observed by MAGIC and VERITAS for short durations, collecting 15 and 10 hours of data, respectively, which led to upper limits consistent with Milagro \citep{Bartko_2008,Kieda_2008}. Soon after, the Tibet Air Shower array confirmed the Very-High-Energy (VHE) extended source detection \citep{Amenomori_2008}. However, ARGO-YBJ later reported a non-detection of MGRO~J2019+37 and reported upper limits for that location. This was attributed to either insufficient exposure to the source or variability of the emission over time \citep{Bartoli_2012}. 

Deep observations of the MGRO~J2019+37 region by VERITAS resolved it into two sources: VER~J2019+368 and VER~J2016+371 \citep{Aliu2014}. VER~J2019+368 is a brighter extended source, which accounts for the bulk of the emission from MGRO~J2019+37, though the origin of the emission remained unknown. VER~J2016+371 is the other source, which is a point-like source to VERITAS. It was suggested that the most likely counterpart of VER~J2016+371 is a Pulsar Wind Nebula (PWN) in the SuperNova Remnant (SNR) CTB 87 because of the co-location of the TeV \gam{}-ray and X-ray emission as well as the luminosity in those energy ranges. 

The MGRO~J2019+37 region overlaps with several SNRs, HII regions, Wolf-Rayet (WR) stars, high-energy \gam-ray ($>$ 100 MeV) sources, and the hard X-ray transient IGR~J20188+3647 (17-30 keV) \citep{igr_sguera,Abdo_2012}. A young energetic radio and \gam{}-ray pulsar, PSR~J2021+3651, \citep{Roberts2002,PSR2021_gamma_ray} and its nebula, PWN~G75.1+0.2, are also in the vicinity \citep{Abdo_2007b}. PSR~J2021+3651 has a period ($P$) of 104 ms, spin-down luminosity ($\dot{E}$) of $3.4\times10^{36}$ erg s$^{-1}$ and a characteristic age ($\tau_c$) of 17.2 kyr \citep{Roberts2002}.  

PSR~J2021+3651 has previously been proposed to be the engine powering the PWN giving rise to the extended TeV emission seen by Milagro, due to its high $\dot{E}$ \citep{saha_and_bhatt}. However, we note that \cite{Paredes_2009} have suggested that PSR~J2021+3651 cannot power the whole MGRO~J2019+37 region alone, and proposed that the star-forming region Sharpless 104 (Sh~2-104) may contribute some fraction of the measured flux. Another proposed scenario for the VHE emission was the winds from WR stars in the young cluster Ber 87, although this model proved challenging to fit to the available data \citep{Bednarek_2007}. 

\textit{Chandra} ACIS-S observations of the region around PSR~J2021+3651 lead to the detection of an X-ray PWN~G75.0+0.1 \citep{Hessels_2004}. The peak of TeV emission is offset by $\sim$20$'$, and the measured size of X-ray PWN is $\leq15'$ \citep{Aliu2014}. Due to the significantly smaller size of the X-ray PWN compared to the size of the TeV emission, it was difficult to draw any conclusion regarding their association. In a recent detailed spectral and morphological study of the X-ray PWN using the data from \textit{Suzaku}-XIS and \textit{XMM-Newton}, there is an indication that X-ray PWN and TeV emission are associated, and the PWN associated with PSR~J2021+3651 is a major contributor to the TeV emission, explaining about 80\% of the emission \citep{Mizuno2017}. The location of the peak TeV \gam-ray emission in VER~J2019+368 is offset from the pulsar location \citep{Aliu2014}, which is a typical behaviour in PWNe powered by pulsars of similar age. This behaviour can be explained if electrons and positrons propagate away from the pulsar and lose energy, leading to energy-dependent morphology and the highest energy \gam-ray emission seen where the high-energy $e^\pm$ originate.

A scenario in which high energy leptons are produced in the PWN powered by PSR~J2021+3651 is supported by the TeV emission of VER~J2019+368 and the X-ray morphology \citep{Aliu2014,Mizuno2017}.
The hard spectral index ($1.75\pm0.3$) measured by VERITAS for VER~J2019+368 in \cite{Aliu2014} resembles Vela X. The morphology of Vela X, the only other PWN system exhibiting double tori, powered by a pulsar similar to PSR~J2021+3651, also favors this as a source of high energy leptons \citep{Hessels_2004,VanEtten2008}.  

The Cygnus region is prominently visible in HAWC Observatory sky-maps. 2HWC~J2019+367 is the source associated with MGRO~J2019+37 and VER~J2019+368 \citep{Abeysekara2017}. Our updated analysis allows measurements of \gam{}-ray flux above 50 TeV, which is critical for observing sources like eHWC J2019+368(one of three HAWC sources that have significant emission above 100 TeV). The emission in this energy regime makes eHWC J2019+368 the site of one of the most energetic particle accelerators in our galaxy \citep{Abeysekara2020}.

We will examine the morphology and spectrum of this source in this work, which combines and extends the prior analyses presented in \citep{brisbois2019,joshi2019}. The study of the energy-dependent morphology will contribute towards establishing the nature of the TeV emission. This is enabled by recent improvements to energy estimation and corresponding sky-maps available for the HAWC data \citep{Abeysekara2019}. Additionally, the high-energy sensitivity will enable us to probe spectral features such as the existence of a spectral softening beyond the VERITAS energy range, which would be expected if the TeV emission originates from inverse Compton (IC) scattering. Significant softening in the spectrum favors a leptonic scenario, because of the suppression of the \gam-ray emission due to Klein-Nishina (KN) effects \citep{Moderksi:2005}.

\section{Data and Methods}
%\subsection{The HAWC Observatory}
The High Altitude Water Cherenkov (HAWC) observatory is an extensive air shower array sensitive to astrophysical \gam{}-ray flux from $\sim$300 GeV to $>$100 TeV. It is located at 18\degr59\arcmin41\arcsec{}N, 97\degr18\arcmin30.6\arcsec{}W on Sierra Negra in Mexico. HAWC consists of 300 Water Cherenkov Detectors (WCDs), which each contain four Photomultiplier Tubes (PMTs). More information on the design and construction of HAWC can be found in \cite{Abeysekara2017}. The HAWC analysis here uses the same ground parameter energy reconstruction technique presented in \cite{Abeysekara2019} with data from June 2015 to July 2018 totaling 1038.8 days of data.

\begin{figure}[htp]
%\plotone{j2019_all_data_viridis.png}
\includegraphics[width=\linewidth]{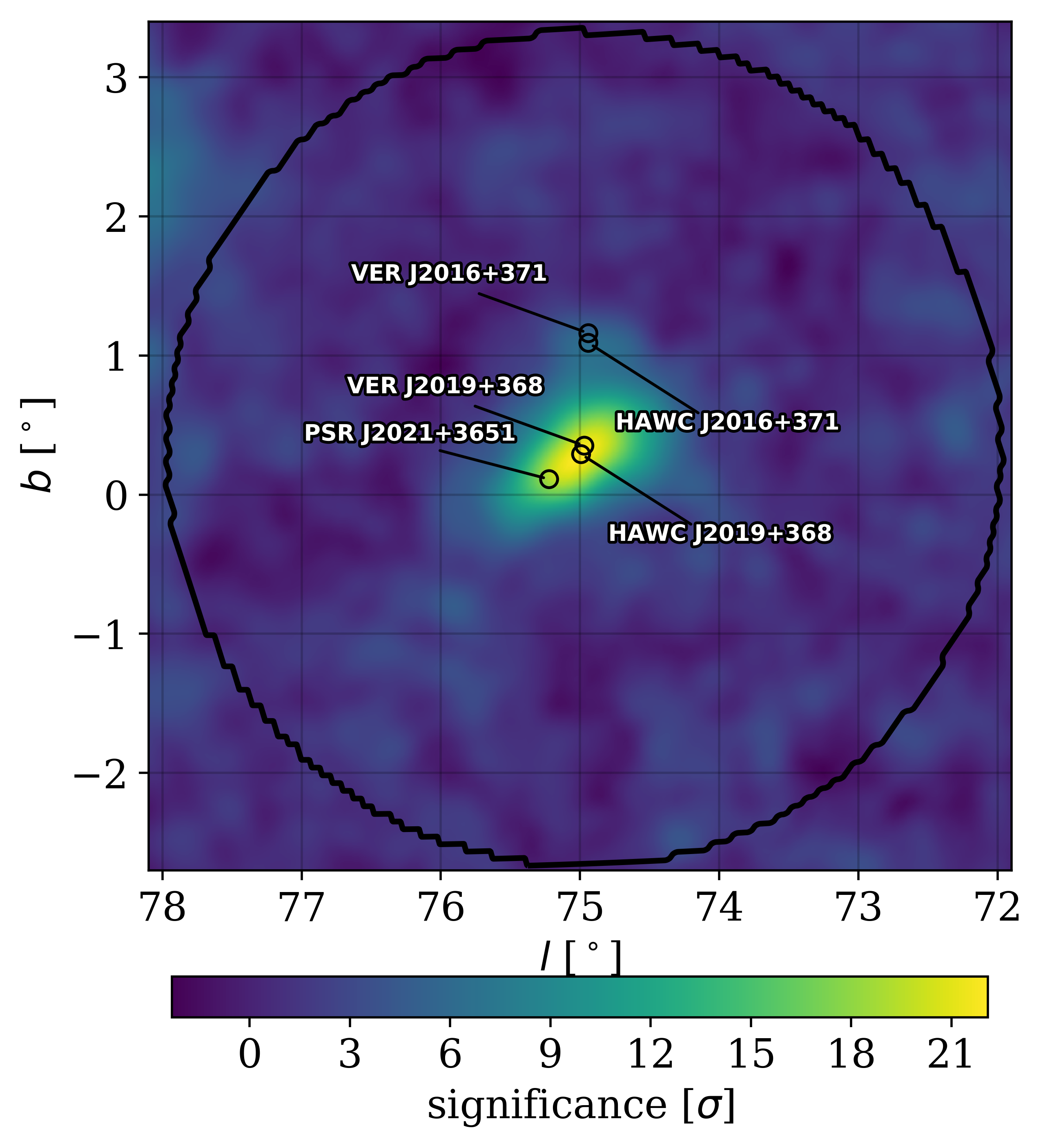}
\caption{Significance map with relevant sources labelled assuming an $\alpha=-2.7$ spectrum. The ROI is indicated by the black circular contour. VERITAS source positions are taken from \cite{Abeysekara2018a}. The position of PSR~J2021+3651 is taken from \cite{Roberts2002}.}
\label{fig:datamap}
\end{figure}

The \gam-ray spectrum and morphology are fit simultaneously using the Multi-Mission Maximum Likelihood (\texttt{3ML}) Framework with the \texttt{hawc\_hal} plugin \citep{threeML2015,hawchal}. The analysis was performed using a Region Of Interest (ROI) with a 3\degr{} radius centered at (l,b)=(75\degr,0.3\degr). This ROI is shown in Figure \ref{fig:datamap}. 

First, the optimal morphological model of the source is determined. This morphology study only considers power law spectral models. Once the best morphological model is found, a search for spectral curvature is performed, using the optimal morphological model found in the first study and again fitting the parameters for the morphology of the HAWC~J2019+368.  

A likelihood ratio (difference in test statistic: $\Delta\mathrm{TS}=~-2\log({L_0}/{L_1})$, following the same convention in \cite{Abeysekara2020}) is performed for nested models to determine which model is preferred. However, when models are not nested, the Bayesian Information Criterion ($\mathrm{BIC}$) is used \citep{Kass1995,Liddle2007}. This is particularly important for comparing exponentially cutoff power laws and log parabolic spectral models, which have the same number of degrees of freedom.

An additional study is performed looking for energy-dependent morphology or a shift in the centroid of emission by making longitudinal profiles oriented along the position angle of the line joining the PSR~J2021+3651 and the best fit location of HAWC~J2019+368. This is similar to that performed in \cite{Aliu2014}, \cite{Aharonian2006a}, and \cite{Abdalla2019}. This study is distinct from the previous morphological/spectral study, and only considers the morphology of the excess counts above the background. This approach allows us to probe the morphological features in the data without assuming a particular model of the morphology and spectrum.  The details and results are described in Section \ref{sec:energy dep morph}.

Finally, using GAMERA \citep{gamera} these results are used to examine the underlying particle distribution under the assumption that they are produced by Inverse Compton (IC) scattering. More details on the modeling and interpretation are discussed in Section \ref{sec:specral modelling} and \ref{sec:discussion}.

\section{Morphological and Spectral Fit}
\label{sec:morphologicalandspectralfit}

\begin{figure}[htp]
%\plotone{single_source_residual.png}
\includegraphics[width=\linewidth]{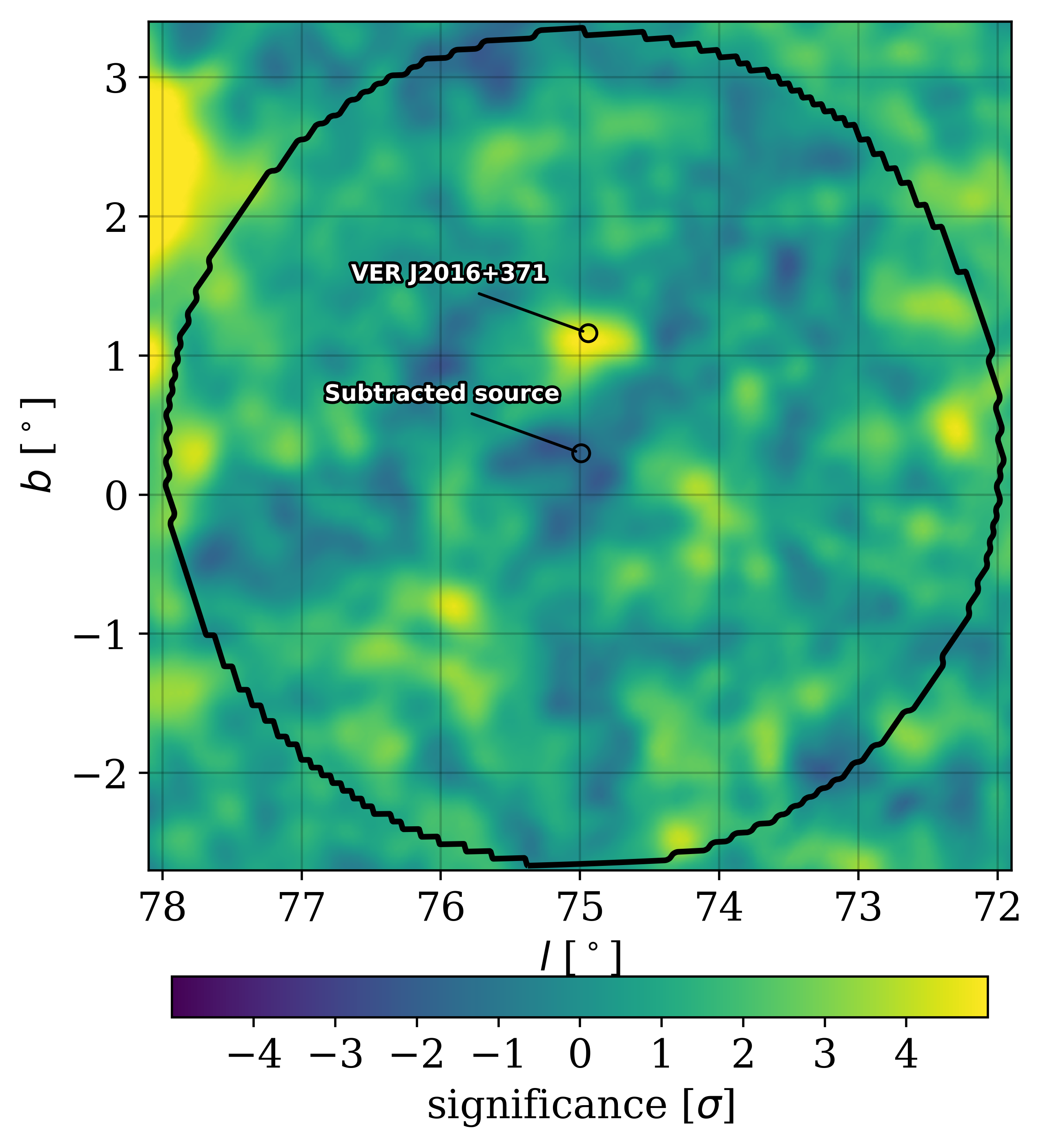}
\caption{Analysis region with a single source subtracted (corresponding to modeling HAWC~J2019+368 as an elliptical Gaussian with a power law spectrum). The maximum significance in the ROI is $\sqrt{\rm{TS}}=5.1$. This new source is called HAWC~J2016+371.}
\label{fig:second_source}
\end{figure}

As shown in Figure \ref{fig:datamap}, this region is dominated by emission near the center of the ROI.  The best-fit model is comprised of two sources: HAWC~J2019+368 and HAWC~J2016+371 on top of a uniform background. The background accounts for emission from extended sources potentially leaking into the ROI, such as one or more extended or diffuse \gam{}-ray sources. At first, when considering a single source morphological hypothesis, a significant excess was seen nearly coincident with VER~J2016+371. Adding a point source, HAWC~J2016+371, at the location of the maximum residual improved the model by $\Delta\mathrm{TS}=39.0$. The need for a second source is also clear from Figure \ref{fig:second_source}, where a $5\sigma$ excess is visible at the location of VER~J2016+371. 

Then, including a uniform background source improved the model consisting of HAWC~J2019+368 and HAWC~J2016+371 by $\Delta\mathrm{TS}=29.3$. This model, consisting of two \gam-ray sources and a background, is then taken as the nominal source model of the region. HAWC~J2019+368 and HAWC~J2016+371 are located at (RA, Dec) = (304.92\degr, 36.76\degr) and (RA, Dec) = (304.1\degr, 37.2\degr), respectively. The model for HAWC~J2019+268 is an elliptical gaussian shape, with parameters for the semimajor axis ($\mathrm{a}$), eccentricity ($\mathrm{e}$), and rotation describing the angle between the line of constant declination and the major axis ($\theta_{\rm {rot}}$). The model corresponds to an ellipse with minor axis $0.12\degr\pm0.02\degr$. The pointing uncertainty on the source positions is $0.10\degr$ \citep{Abeysekara2017a}.

\begin{table*}
\begin{center}
\begin{tabular}{ccc} 
\tableline
Source Name & Spectral Parameters & Morphology\\
\hline \hline
HAWC~J2019+368 
&
 \begin{tabular}{@{}c@{}} 
$\phi_{10\,\mathrm{TeV}}=4.05\pm0.26 \times10^{-14}$  \\ 
$\alpha=-2.02\pm0.06$ \\ 
$\beta=-0.29\pm0.05$ \\ \end{tabular}

& 
\begin{tabular}{@{}c@{}} 
$\mathrm{a}=0.368\degr\pm0.021$\degr  \\ 
$\mathrm{e}=0.943\pm0.017$ \\ 
$\theta_{\rm {rot}}=21.7\degr\pm2.5\degr$ \\ 
\end{tabular} \\ \hline

HAWC~J2016+371 & \begin{tabular}{@{}c@{}} $\phi_{10\,\mathrm{TeV}}=2.6^{+0.7}_{-0.5}\times10^{-15}$  \\ $\alpha=-2.32\pm0.18$ \\  \end{tabular}
 & Point Source \\ \hline
Background &  \begin{tabular}{@{}c@{}} $\phi_{10\,\mathrm{TeV}}=8.2^{+1.5}_{-1.3}\times10^{-14}$  \\ $\alpha=-2.75\pm0.08$ \\  \end{tabular} 
& Uniform over ROI \\
 \hline
\end{tabular}
\caption{Description of model parameters assuming HAWC~J2019+368 has a log parabolic spectrum. $\phi_{10\,\mathrm{TeV}}$ is the flux normalization at 10 TeV in units of $\mathrm{TeV}^{-1}\,\mathrm{cm}^{-2}\,\mathrm{s}^{-1}$. Reported uncertainties are statistical.}
\end{center}
\label{tab:spectrum_lp}
\end{table*}

\begin{table*}
\begin{center}
\begin{tabular}{ccc} 
\tableline
Source Name & Spectral Parameters & Morphology\\
\hline \hline
HAWC~J2019+368 
&
 \begin{tabular}{@{}c@{}} 
$\phi_{10\,\mathrm{TeV}}=4.8^{+0.5}_{-0.4} \times10^{-14}$  \\ 
$\alpha=-1.67\pm0.10$ \\ 
$E_{\rm {cut}}=37^{+8}_{-7}$ TeV \\ \end{tabular}

& 
\begin{tabular}{@{}c@{}} 
$\mathrm{a}=0.358\degr\pm0.022$\degr  \\ 
$\mathrm{e}=0.953\pm0.017$ \\ 
$\theta_{\rm {rot}}=21.9\degr\pm2.6\degr$ \\ 
\end{tabular} \\\hline

HAWC~J2016+371 & \begin{tabular}{@{}c@{}} $\phi_{10\,\mathrm{TeV}}=2.9^{+0.7}_{-0.6}\times10^{-15}$  \\ $\alpha=-2.28\pm0.17$ \\  \end{tabular}
 & Point Source \\ \hline
Background &  \begin{tabular}{@{}c@{}} $\phi_{10\,\mathrm{TeV}}=8.1^{+1.5}_{-1.3}\times10^{-14}$  \\ $\alpha=-2.74\pm0.09$ \\  \end{tabular} 
& Uniform over ROI \\
 \hline
\end{tabular}
\caption{Description of model parameters assuming HAWC~J2019+368 has an exponentially cutoff power law spectrum. $\phi_{10\,\mathrm{TeV}}$ is the flux normalization at 10 TeV in units of $\mathrm{TeV}^{-1}\,\mathrm{cm}^{-2}\,\mathrm{s}^{-1}$. Reported uncertainties are statistical.}
\end{center}
\label{tab:spectrum_cpl}
\end{table*}

\begin{figure}[htp]
\includegraphics[width=\linewidth]{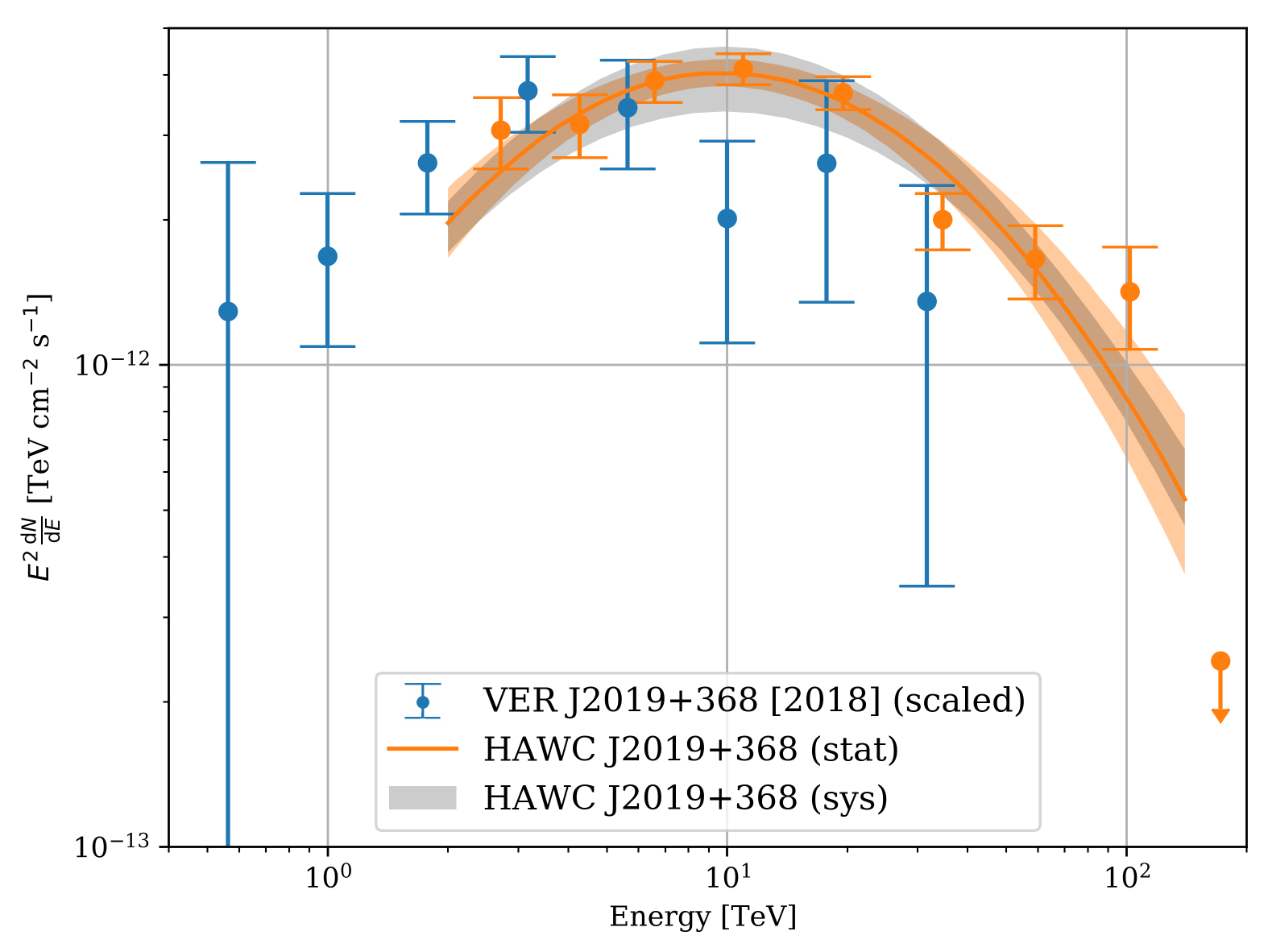}
\caption{Spectral energy distribution of HAWC~J2019+368. VERITAS data points are taken from \cite{Abeysekara2018a} after the scaling detailed in Section \ref{sec:scaling}. Upper limits are reported for energy bins whose $TS$ is less than 4.\label{fig:sedJ2019}}
\end{figure}

\begin{figure}[htp]
\includegraphics[width=\linewidth]{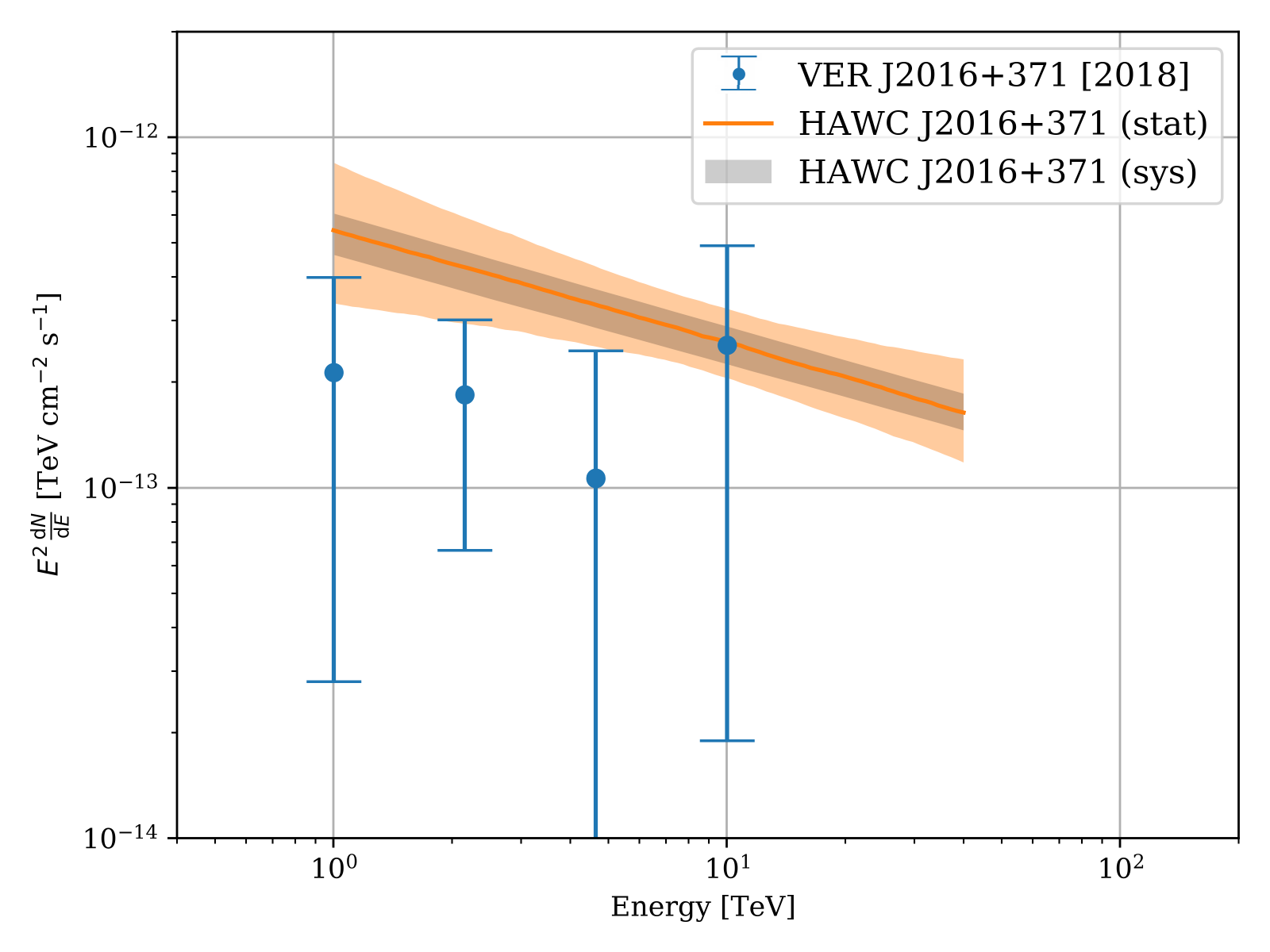}
\caption{Spectral energy distribution of HAWC~J2016+371. VERITAS data points are taken from \cite{Abeysekara2018a}. No scaling of VERITAS data points was performed as discussed in Section \ref{sec:scaling}.}
\label{fig:sedJ2016}
\end{figure}

A study looking for curvature in the spectrum of HAWC~J2019+368 is then performed. A log parabola (power law with exponential cutoff) spectral assumption for HAWC~J2019+368 is significantly preferred over a pure power law model by $\Delta\mathrm{TS}=69.0$ ($58.9$). Using $\mathrm{BIC}$ to find the preferred model between the (non-nested) curved spectral models, a log parabola spectrum is preferred with a $\Delta\mathrm{BIC}=10.1$. The best fit parameters for the reported log parabola spectrum are shown in Table \ref{tab:spectrum_lp}. The best fit parameters for an exponentially cutoff power law are shown in Table \ref{tab:spectrum_cpl}. The spectral energy distribution of HAWC~J2019+368 is shown in Figure \ref{fig:sedJ2019}, with the VERITAS data points from \cite{Abeysekara2018a} scaled by 2.7 according to the procedure described in Section \ref{sec:scaling}. The HAWC~J2016+371 spectral energy distribution is shown in Figure \ref{fig:sedJ2016}, with the corresponding data points from  \cite{Abeysekara2018a}. HAWC~J2016+371 was not significantly detected in any individual energy bin, and therefore only the fit is reported here.

The systematic uncertainty bands shown in Figures \ref{fig:sedJ2019} and \ref{fig:sedJ2016} are produced by fitting the best model described above while varying the simulated model of HAWC. The same procedure and detector response files were used to produce the systematic uncertainty band as in \cite{Abeysekara2019}. The plotted energy range is determined by adding a step function cutoff to the nominal model, and stopping when the $\Delta\mathrm{TS}$ equals one between the nominal model and the nominal model with the cutoff. This procedure is done twice ---once at the higher energies and once at the lower energies--- to determine the energy range over which this model describes the source.

\subsection{Comparison to VERITAS measurements} \label{sec:scaling}

\begin{figure}[ht!]
%\plotone{integration_regions.png}
\includegraphics[width=\linewidth]{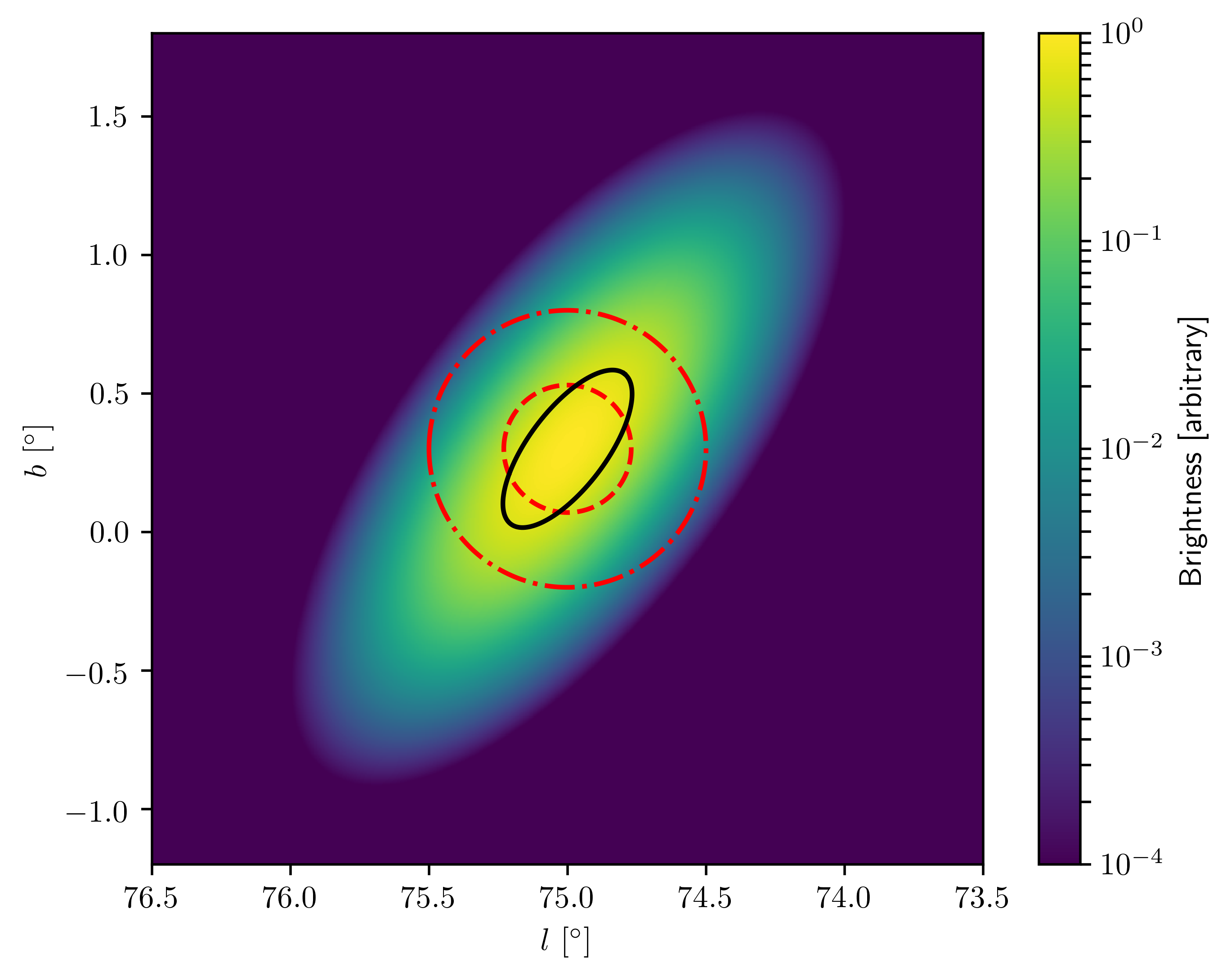}
\caption{VERITAS Morphology map of VER~J2019+368 (taken from \cite{Abeysekara2018a}). The smaller red contour is the 0.23\degr{} extraction region from which the spectrum reported in \cite{Abeysekara2018a} is taken. The larger red dashed contour is the 0.5\degr{} from which the spectrum reported in \cite{Aliu2014} is taken. The black elliptical contour is the 1$\sigma$ contour for the morphology reported in this work for HAWC~J2019+368. The color scale is arbitrary and scaled such that the maximal pixel value is 1. \label{fig:scale_veritas}}
\end{figure}

From their studies published in 2014 and 2018, the VERITAS collaboration extracted a spectrum from a smaller region than the reported morphology of the source \citep{Aliu2014,Abeysekara2018a}. This is different from HAWC, where the morphology and spectrum are fit simultaneously in the ROI. This leads to a predictable systematic offset in the reported flux between the two instruments when two conditions are met: the measured morphology by VERITAS is larger than the extraction region, and the morphologies measured by HAWC and VERITAS are similar. This is not as significant an issue for sources that appear point-like for both instruments such as VER~J2016+371 \citep{Abeysekara2018a} and HAWC~J2016+371 (this work) or the Crab Nebula \citep{Abeysekara2019,Meagher2015}.

To account for this, we divide the flux reported by VERITAS by the fraction of the reported morphology contained within the extraction region. This has the effect of scaling the flux up to the morphology by a factor of $2.71$, under the assumption that the spectrum is the same for the entire source extent. In Figure \ref{fig:scale_veritas} one can see that a substantial fraction of the source reported in \cite{Aliu2014} and \cite{Abeysekara2018a} is not contained by the extraction region used to report the spectrum. 

\section{Energy Dependent morphology}
\label{sec:energy dep morph}
To study the potential energy-dependent morphology of HAWC~J2019+368, we examine the longitudinal profile of the data. The profile of the excess counts over the background is obtained within a region of the sky in different energy bands. 
The quarter decade energy bins were combined into four energy bands from 0.3-1.7 TeV, 1.7-10 TeV, 10-56 TeV, and 56-316 TeV to perform the study. A selection of the 2D bins ( as described in \cite{Abeysekara2019}) is done in each energy band (fraction hit bins greater than 2, 5, 7, and 9 for the energy bands 1, 2, 3, and 4, respectively) so that the best Point Spread Function (PSF) possible is retained without significantly losing statistics in each bin.

The region taken for the profiles is centered at the location of the PSR~J2021+3651. The orientation of the region is along the position angle of $-77.90$\degr, which is the line joining the PSR~J2021+3651 and HAWC~J2019+368 positions. This orientation was chosen to test the hypothesis that the peak of the TeV emission would move to the pulsar location with increasing energy together with the change in the size of the emission region.

\begin{figure}[ht!]
\includegraphics[width=\linewidth]{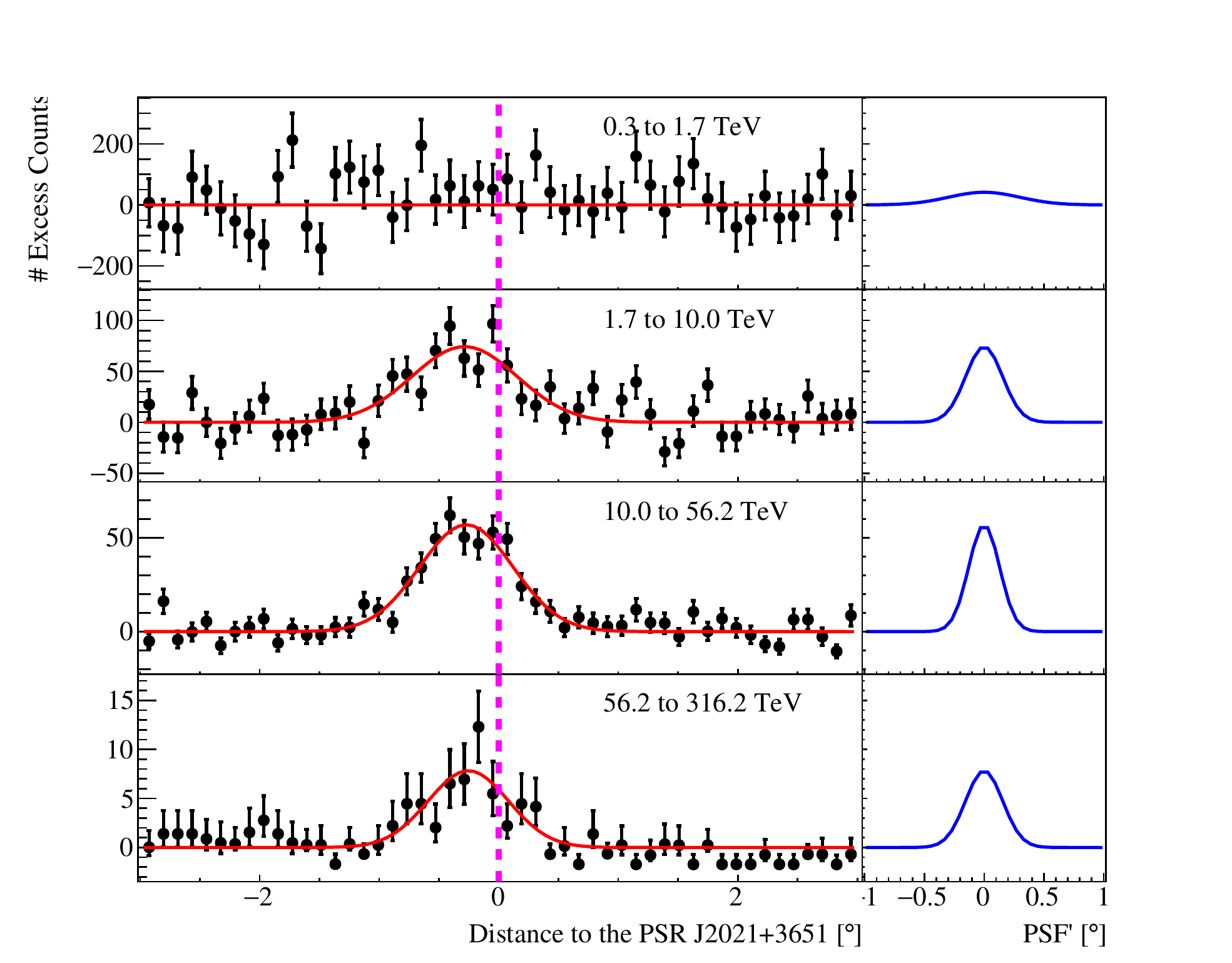}
\caption{The longitudinal profiles of the excess counts maps for HAWC~J2019+368 in different energy bands. The corresponding PSF$^{\prime}$ is shown on the right of each plot in blue (see Section \ref{sec:energy dep morph}). The profiles are fitted with a Gaussian shown in red. The location of the pulsar PSR~J2021+3651 is shown with the magenta dashed line.}
\label{fig:slice_profile_plot_j2019}
\end{figure}

The profiles are 0.7\degr $\times$ 6\degr{} on the sky with 50 bins in the four reconstructed energy bands and are shown in the left panel of Figure \ref{fig:slice_profile_plot_j2019}.
The width (0.7\degr{}) was chosen to avoid possible contamination from HAWC~J2016+371 in the region of HAWC~J2019+368.  There was no significant emission observed in the first energy band, which might be due to the hard spectrum of HAWC~J2019+368. The measured spectrum for HAWC~J2019+368 also shows that the first significant flux point is at $>2$ TeV (see Figure \ref{fig:sedJ2019}), which is in agreement with the upper bound of the first energy band at 1.77 TeV.  We see significant excess in energy bands 2, 3, and 4, which corresponds to the energy range above 1.77 TeV. The excess profiles were fitted with a Gaussian function and examined for size and location of the centroid of emission relative to PSR~J2021+3651. 

To estimate the size of the PSF in each longitudinal profile, we simulated a point source  at the location of PSR~J2021+3651 with a power law of spectral index 2.2. Similarly, excess count profiles are obtained and fitted with a Gaussian for this simulated point source. This $1\sigma$ width of the fitted Gaussian is hereafter referred to as PSF$^{\prime}$ for the purposes of this study. The PSF$^{\prime}$ obtained for different energy bands are shown in the right panel of Figure \ref{fig:slice_profile_plot_j2019}.

\begin{figure}[ht!]
\includegraphics[width=\linewidth]{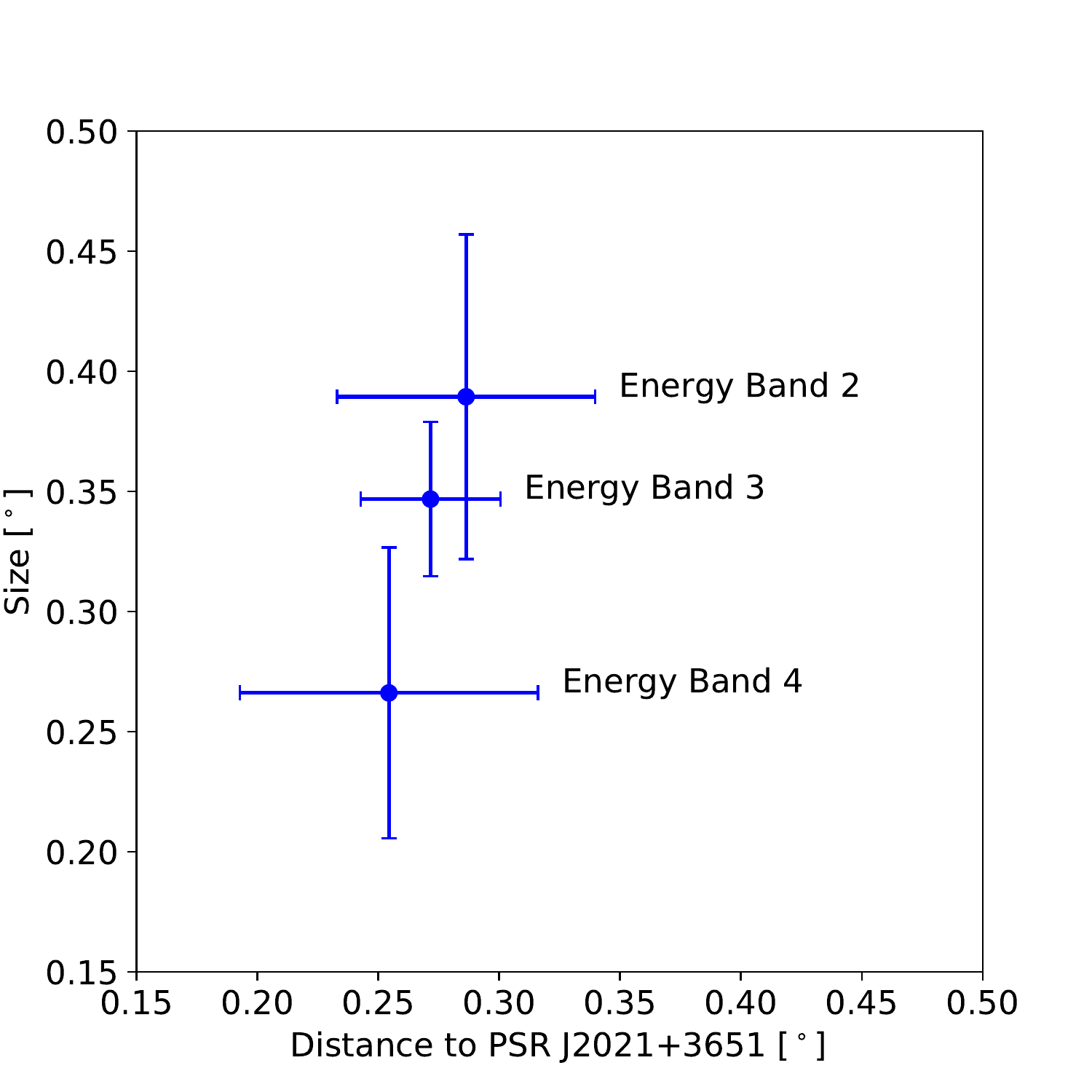}
\caption{The size vs. distance with respect to the pulsar PSR~J2021+3651 in different energy bands for HAWC~J2019+368. Size is defined as the $\sqrt{\sigma_{\rm Fit}^{2} - (\sigma_{\rm PSF^{\prime}} + 0.05)^{2}}$, where $\sigma_{\rm Fit}$ and $\sigma_{\rm PSF^{\prime}}$ is the $1\sigma$ width of the fitted Gaussian as shown in Figure \ref{fig:slice_profile_plot_j2019}. The PSF$^{\prime}$ is corrected for the systematic difference of 0.05$^\circ$ between data and simulations (see Section \ref{sec:energy dep morph}). The mean position of the fitted Gaussian is used to calculate the distance to the pulsar location.}
\label{fig:size_vs_shift_j2019_plot}
\end{figure}

In Figure \ref{fig:size_vs_shift_j2019_plot} we show the size and distance of the peak TeV emission with respect to the pulsar location in energy bands 2, 3, and 4 using the excess count profiles. The size was obtained by subtracting the $1\sigma$ width of the PSF$^{\prime}$ in quadrature from the $1\sigma$ fit size of the excess profiles in each energy band. The $1\sigma$ width of the PSF$^{\prime}$ was corrected by $0.05^\circ$ to account for a systematic difference of 5\% between simulations and data \citep{Abeysekara2019}. The distance to the pulsar was calculated from the mean position of the excess profile fit. There is a very mild indication of the decrease in size with energy bands of increasing energy. However, the uncertainty on the measured size is large and there is no significant shift towards the pulsar position with increasing energy.

\section{Spectral Modeling}
\label{sec:specral modelling}
Amongst the different scenarios that can explain the multi-wavelength emission from HAWC~J2019+368, the observed spatial association between the X-ray and TeV emission, 
as well as the mild indication that the size of the TeV emission shrinks with increasing energy, seem to point towards a leptonic origin. Therefore we limit this analysis to a leptonic scenario. Exploring a hadronic or a lepto-hadronic origin of the TeV emission and their comparison is beyond the scope of this work. 

In this section, we present a model assuming a leptonic scenario where the emission is originated by a PWN powered by PSR~J2021+3651. The observed period ($P$) and its time derivative ($\dot{P}$) are 104 ms and $9.57 \times 10^{-14}$ s s$^{-1}$, respectively, for PSR~J2021+3651 \citep{Roberts2002}. The characteristic age, $\tau_c$, of the pulsar is 17.2 kyr (see Equation \ref{eqn:age_pulsar}). It is worthwhile to note that the $\tau_c$ is only an accurate measure of the pulsar age, 
\begin{equation}
\tau = \frac{P}{(n-1) \dot P} \left[ 1-\left( \frac{P_0}{P} \right)^{n-1} \right],
\label{eqn:age_pulsar}
\end{equation}
when the assumption of braking index $n=3$ and the birth period $P_0 \ll P$ is true \citep{Gaensler2006}. For the distance to the pulsar, the dispersion-based measurement is quite large, $\geq$ 10~kpc \citep{Roberts2002}, which conflicts with distance derived considering the X-ray measurements \citep{VanEtten2008} of $\sim4$~kpc. The recent study of \textit{Chandra} archival X-ray data by \cite{Kirichenko2015}, taking into account an extinction-distance relation using the red-clump stars as standard candles in the line of sight, suggests a distance of $\sim\!\!1.8$~kpc. Therefore in this work, we will be using the distance of 1.8~kpc. 

In the following, we will describe the model in Section~\ref{sec:model} and then move to two scenarios where the model will be used to describe the observed X-ray and TeV emission together. In the first scenario, designated the one zone model (see Section~\ref{one zone model}), the observed X-ray and the whole range of TeV emission are described together. However, it is not ideal to model the X-ray emission between 2 and 10 keV (hard X-ray emission) together with TeV emission between $\sim\!\!1$ TeV and at least $\sim\!\!100$ TeV \gam-ray energy because different components of the parent electron population (hereafter meaning both electrons and positrons) are responsible for the emission measured at X-ray and TeV energies. The size of the X-ray emission, $\sim\!\!4'\times10'$ in full width at half maximum \citep{Mizuno2017}, is very small compared to the HAWC TeV emission of size $\sim\!\!20'\times50'$ ($1\sigma$ containment, see Table~\ref{tab:spectrum_lp}). The component of the electron population emitting hard X-rays at 2-10 keV must be the one emitting the highest energy \gam{}-rays at several tens of TeV and must be of higher energy than the one responsible for the extended TeV emission. This is because lower energy electrons have longer cooling timescales and can still produce TeV emission by traveling farther.

In the second scenario, the two zone model (see Section \ref{two zone model}), the X-ray emission is compared with the highest energy TeV emission. Only the recently injected electrons will be used to describe the X-ray emission together with the whole history of injected electrons to describe the TeV emission. The possibility of undetected X-ray emission, comparable to the size of the TeV emission region will be taken into account to describe the whole history of the system. %The two zone model scenario is preferred for the interpretation over the one zone model. 

\subsection{Model Considerations}
\label{sec:model}
The pulsar braking index is assumed to be $n = 3$. The cut-off energy ($E_{\rm max}$), the birth period ($P_0$), and the conversion efficiency ($\epsilon$) are treated as unknown parameters to describe the injected particle spectrum.  The true age ($\tau$) of the system is calculated as a function of $P_0$ using Equation \ref{eqn:age_pulsar} instead of using the characteristic age ($\tau_c$). 

We used the GAMERA software to perform the spectral modelling \citep{gamera}. GAMERA can define a time-dependent model of relativistic electrons, including injection and cooling, and output the associated photon emission given a set of radiation fields. We consider a broken power-law with an exponential cut-off for the spectrum of the injected electrons. The break in the injected electron spectrum is believed to be due to two different electron populations (wind and radio) which dominate below and above the break energy ($E_b$) respectively. The value of $E_b$ and the injection spectral indices of the radio and wind populations are set to be their typical values of 0.1 TeV, $-1.5$, and $-2.0$ \citep{Meyer_crab,Atoyan_Aharonian_2006}. A cut-off is treated as a free parameter because the injected particles (e$^{\pm}$) can only be accelerated up to a certain maximum energy ($E_{\rm max}$) \citep{Nonthermal_rad_pwn, Gaensler2006}. The radiation fields used to calculate the IC emission are calculated from the model presented in \cite{radiation_model_tuffs} at the location of PSR~J2021+3651. 

The time evolution of the parameters $B$, $\dot{E}$, and $P$ are specified following \cite{Gaensler2006}. The spin-down time scale ($\tau_0$) of the system is an important input for defining the time evolution of electrons in the system \citep{Gaensler2006,emax_larmor_radius_pwn}. The spin-down timescale can be expressed as a function of $P_0$ as $\tau_0 = P_0^{2}P^{-1}/{(2\dot{P})}$ \citep{Gaensler2006}. This allows us to choose the birth period of the pulsar that gives the best fit to the data. The nebular B-field is given by $B(t)= {B_0}[1 + (t/\tau_0)^{0.5}]^{-1}$.
The normalization of the injected particle spectrum is determined by $\epsilon\times\dot{E}(t)$, the fraction of the spin-down luminosity going into the production of electrons. The function $\dot{E}(t)$ is defined as $\dot{E}(t)= \dot E_0\left( 1+t/\tau_0 \right)^{-2}$. Lastly, the pulsar slows down its rotation according to $P(t)= P_0\left(1 + t/{\tau_0}\right)^{0.5}$.

\begin{figure}[ht!]
\includegraphics[width=\linewidth]{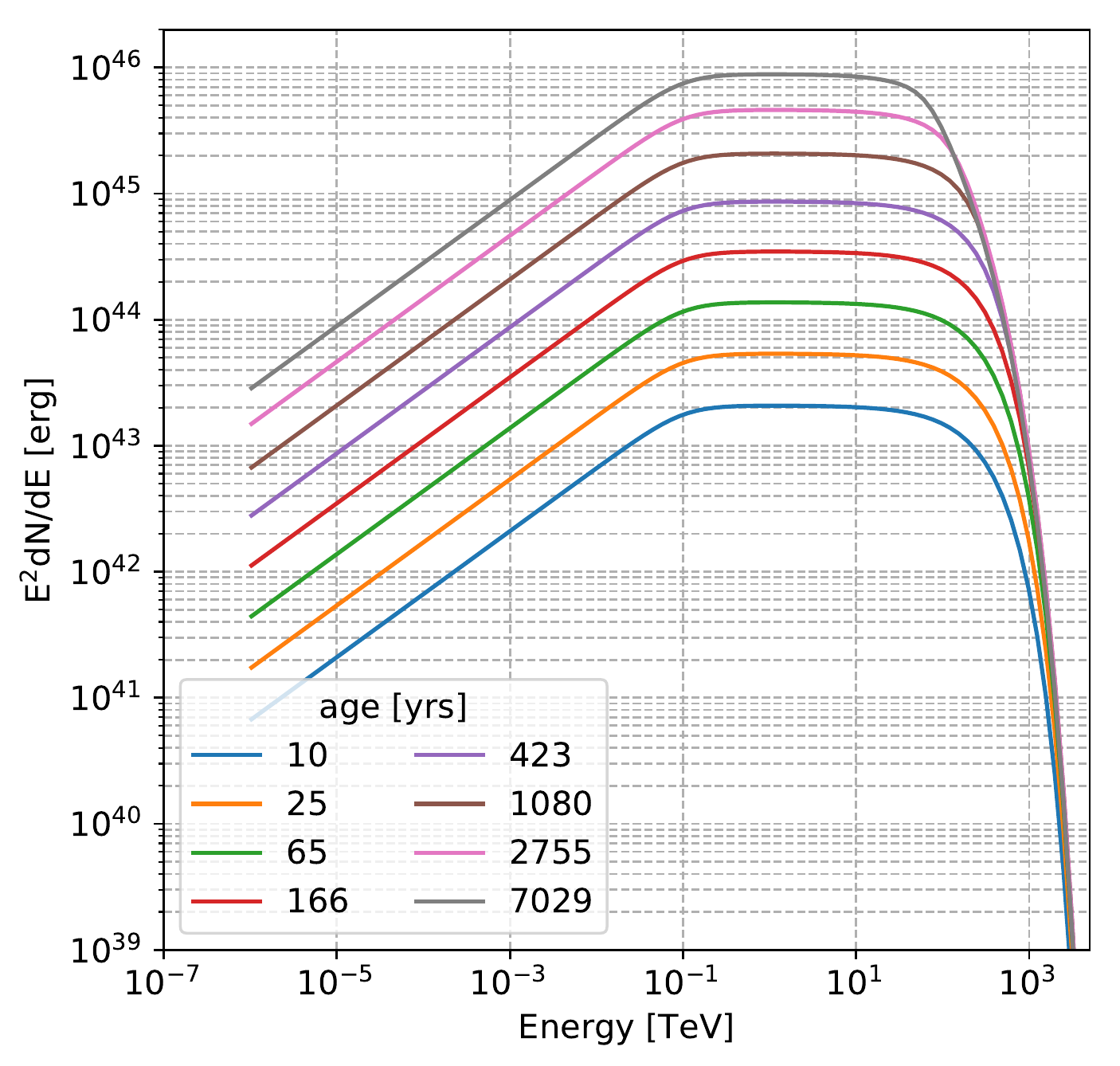}
\caption{Time evolution of the particle e$^{\pm}$ spectrum for the two-zone model described in Section \ref{two zone model}. Ages are equal steps of 0.2 $\log_{10}(t$ years$^{-1})$.}
\label{fig:evolved_particle_spectrum}
\end{figure}
This way, our data constrains the birth period and also estimates the true age of the system. The injected particle spectrum for given values of assumed and unknown parameters is then evolved in the presence of magnetic and radiation fields to calculate the evolved particle spectrum as shown in Figure \ref{fig:evolved_particle_spectrum}. The evolved particle spectrum is used to calculate the emitted photon spectrum, which is then matched to the observed X-ray and TeV emission. 

\subsection{One Zone Model}
\label{one zone model}
\begin{figure}[ht!]
\includegraphics[width=\linewidth]{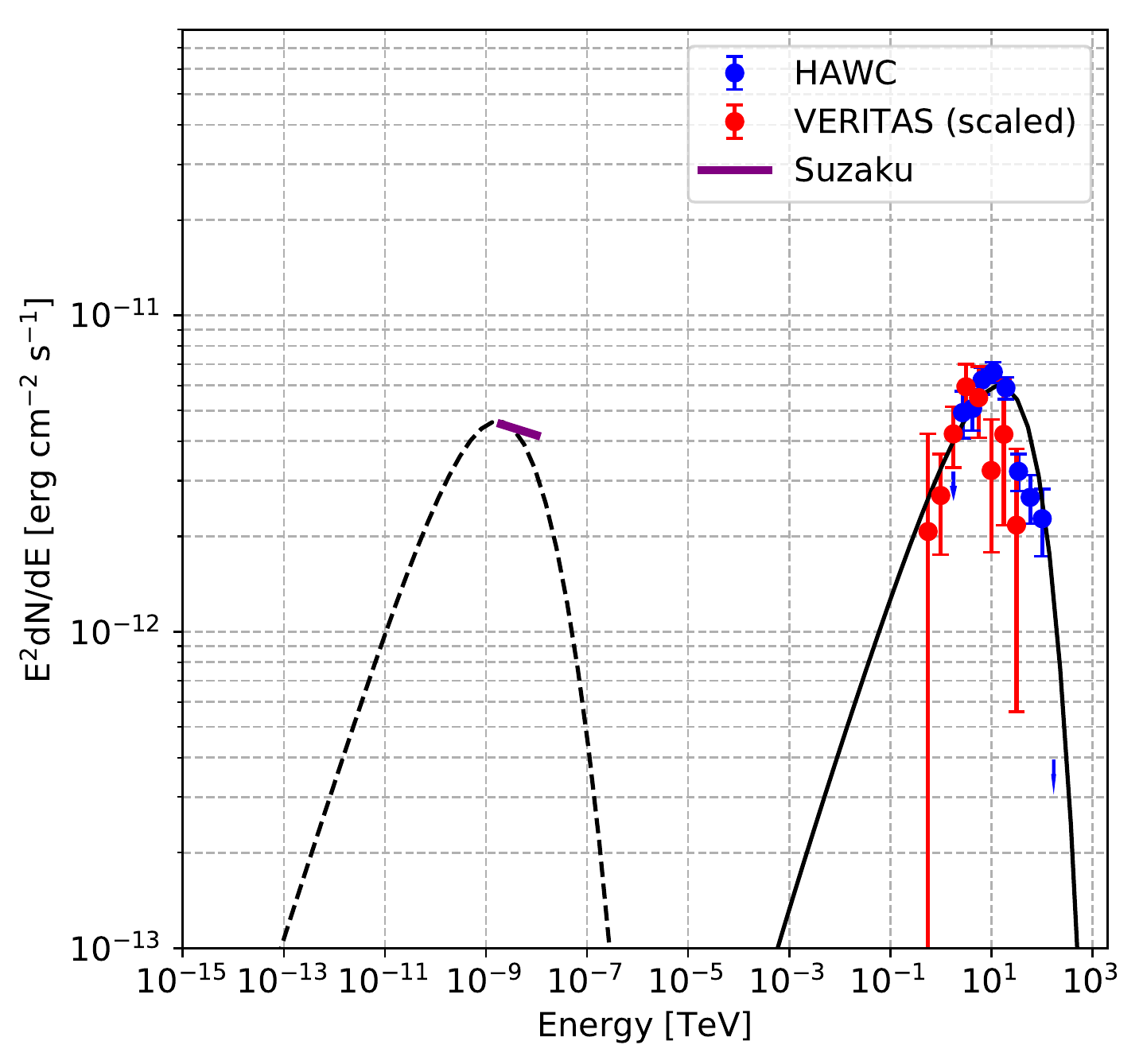}
\caption{One zone model SED for synchrotron and IC are shown with dashed and solid lines respectively. The X-ray SED obtained by \textit{Suzaku} and \gam-ray SEDs for HAWC and VERITAS are shown (same as in Figure \ref{fig:sedJ2019}). The model SED is tuned manually to provide an approximate match to the X-ray and \gam-ray emission in the whole energy range. The present-day B-field is 2 $\,\mu$G}
\label{fig:one_zone_model}
\end{figure}

The emitted photon spectrum of the one zone model is shown in Figure \ref{fig:one_zone_model}. The model synchrotron (dashed lines) and IC (solid lines) emission are manually tuned to match the observed X-ray and \gam-ray emission. It can be seen that the model synchrotron and IC emission are reasonably consistent with the observed X-ray and \gam-ray data SED. 

The estimated value of the present day magnetic field is 2 $\mu$G, which is low (relative to the ISM) and depends on the relative normalization of the peak of the synchrotron and IC emission. The estimated value for $P_0$ of $80$ ms corresponds to an age of $\sim$7 kyr for the system. The required conversion efficiency ($\epsilon$) of pulsar spin-down power to particle injection is $5\%$, with the estimated cut-off energy ($E_{\rm max}$) of the injected particle spectrum being 300 TeV.  However, there are certain limitations of the outcomes of the one zone model mainly pertaining to the obtained B-field of 2 $\mu$G.  The estimated value of the B-field is very low in comparison to all expectations for a PWN and to the mean B-field of the local ISM ($\sim3\mu$G). This model would predict identical X-ray and \gam-ray morphology, which is in contradiction with the observations, as the observed size of the X-ray PWN is significantly smaller than that of the TeV emission.

\subsection{Two Zone Model}
\label{two zone model}
The low magnetic field derived in the one-zone model indicates that describing the TeV and hard X-ray (2-10 keV) emission together using the whole electron population is not optimal. The limitations of the one zone model can be resolved by considering that the electrons producing hard X-rays are only a subset of the total electron population emitting \gam-rays. The energy of an electron ($E_e$, as given in \cite{Mizuno2017}), emitting synchrotron radiation with mean energy $E_{\rm X-ray}$, due to a magnetic field $B$, is given by 
\begin{equation}
E_e \simeq 132 \text{ TeV} \left( \frac{E_{\rm X-ray}}{\text{1 keV}}\right)^{0.5} \left( \frac{B}{3 \mu {\rm G}}\right)^{-0.5}.
\label{eqn:el_en_xray}
\end{equation}

Assuming a minimum magnetic field strength of 3$\,\mu$G (equal to the local ISM) the corresponding parent electron population component should be of energy $\sim$300 TeV for 5 keV X-rays \citep{Mizuno2017}.

Given the radiation fields from the model at 300 TeV electron energies, the most relevant radiation field is the Cosmic Microwave Background (CMB), with a magnetic field strength of $\sim$3$\,\mu$G. The relevant cooling timescales are of the order of $\sim$2 kyr using  

\begin{equation}
\tau_{\rm cool} \approx 3.1 \times 10^5 \left( \frac{U}{\text{eV} \text{cm}^{-3}}\right)^{-1} \left( \frac{E_e}{\text{TeV}}\right)^{-1} \text{yr},
\label{eqn:cooling}
\end{equation}

where the total energy density, $U$m is defined as $U = f_{\rm KN} U_{\rm ph} + 0.025(B/{\mu}{\rm G})^2$ eV cm$^{-3}$ with respect the radiation field energy density $U_{\rm ph}$), the magnetic field  ($B$), and $f_{\rm KN}$, the normalization factor to take into the suppression of IC losses due to KN effects \citep{Jim_tev_astro}.
The normalization factor can be calculated as:

\begin{equation}
f_{\rm KN} \approx \left(1 + 15 \frac{E_e}{\text{TeV}} \frac{E_T}{\text{eV}}\right)^{-1.5},
\end{equation}
where $E_T$ is the target photon energy \citep {Jim_tev_astro}. 
%\textcolor{purple}{Chad: I tried to make this sentence more clear. does it make sense? $\rightarrow$} 
We manually tune the B-field to match the observed normalization of X-ray emission such that the electrons currently producing hard X-rays are as much as 2 kyr old, resulting in a required present day B-field of 3.5 $\mu$G. 
\begin{figure}[ht!]
\includegraphics[width=\linewidth]{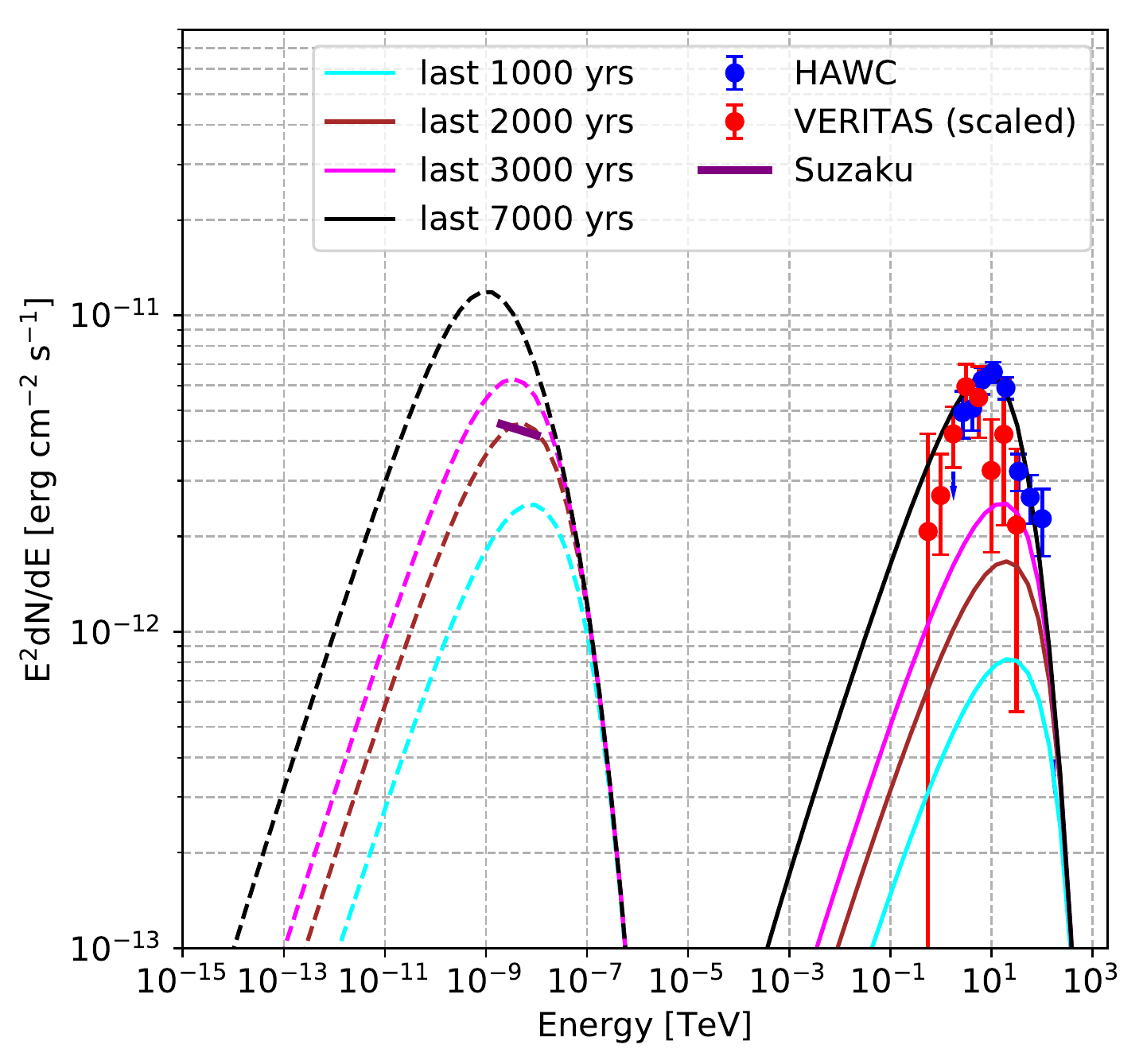}
\caption{Two-zone model SED for synchrotron and IC are shown with dashed and solid lines, respectively. Model SEDs due to electron population injected in last 1, 2, and 3 kyr are also drawn together with the total time-integrated emission. The X-ray SED obtained by \textit{Suzaku} \citep{Mizuno2017} and \gam-ray SEDs for HAWC and VERITAS are shown (same as in Figure \ref{fig:sedJ2019}). The model SED is tuned manually to provide an approximate match to the X-ray emission by electron population injected in last 2 kyr, and TeV emission in the whole energy range by the total injection history ($\sim$7 kyr). The required present-day B-field is 3.5 $\mu$G.} 
\label{fig:two_zone_model}
\end{figure}
To match the model synchrotron and IC peak to the observed X-ray and TeV emission peak, a birth period of 80 ms is required, therefore the true age of the system for the two-zone model agrees with the one-zone model with a value of $\sim$7 kyr. However, due to the increased B-field, the conversion efficiency is marginally increased to a value of 6.5\% in this case. The cut-off energy in the injected particle spectrum remains the same.

In Figure \ref{fig:two_zone_model}, the model SED lines for electrons injected in the last 1, 2 and 3 kyr are  shown together with the total injection history of $\sim$7 kyr.  In this case, the total synchrotron emission overshoots the observed X-ray SED by a factor of 4-5 at 2 keV. This may be due to the smaller region on the sky observed by the X-ray observations compared to the TeV observations, and result in some of the emission being unmeasured. It is also mentioned by \cite{Mizuno2017} that the observed X-ray emission should be taken as the lower limit. 

The X-ray emission seen by \textit{Suzaku} is $\sim\!\!4'\times10'$ in full width at half maximum \citep{Mizuno2017} while the \gam{}-ray emission is about $\sim\!\!20'\times50'$ using the same measure, meaning the TeV emission seen by HAWC is a factor of 25 larger in solid angle compared to the \textit{Suzaku} observations. The dimmest reported flux from the \textit{Suzaku} observations is $\sim\!\!1.7\times10^{-13}$ erg cm$^{-2}$ s$^{-1}$ in a $5'\times3'$ region \citep{Mizuno2017}. By scaling this flux in X-rays from the smaller observed region to the HAWC emission region, one would get a factor of 2-3 higher in measured X-ray flux.

The peak of the X-ray SED at 2 keV is about $\sim\!\!4\times10^{-12}$ erg cm$^{-2}$ s$^{-1}$, therefore by taking into account the expected missing emission it would be around $\sim\!\!1\times10^{-11}$ erg cm$^{-2}$ s$^{-1}$. On Figure \ref{fig:two_zone_model} the present day synchrotron peak is near the value of $\sim\!\!10^{-11}$ erg cm$^{-2}$ s$^{-1}$, whereas the observed peak is coincides with the most recent $\sim\!\!2$ kyr of emission. The observed \gam{}-ray emission can be compared to the model emission, where it can be seen that the model's present-day IC emission is reasonably consistent with the observed TeV emission. 

\section{Discussion}
\label{sec:discussion}
To examine the obtained birth period ($P_0$) of 80 ms and true age ($\tau$) of $\sim\!\!7$ kyr, one can compare HAWC~J2019+368 with the observed behaviour of HESS J1825-137. Both of these sources are bright extended TeV sources with similar associated pulsars. The associated pulsar of HESS J1825-137 is PSR B1823-13. The comparison of the pulsar properties of PSR~J2021-3651 and PSR B1823-13 is shown in Table \ref{tab:pulsar_properties_comparison}.
\begin{table*}
\begin{center}
\begin{tabular}{l c c}
\hline 
Source  & HAWC~J2019+368 & HESS J1825-137\\
\hline \hline   
Associated pulsar & PSR~J2021+3651 & PSR B1823-13\\
Present period ($P$) & 104 ms & 102 ms\\
$\dot{P}$ & $9.57\times10^{-14}$ s s$^{-1}$ & $7.5\times10^{-14}$ s s$^{-1}$ \\
$\dot{E}$ & $3.4\times10^{36}$ erg s$^{-1}$ & $2.8\times10^{36}$ erg s$^{-1}$ \\
Characteristic Age ($\tau_c$) & 17.2 kyr & 21.4 kyr \\
Distance ($d$) & 1.8 kpc & 4 kpc \\
\hline
\end{tabular}
\end{center}
\caption{Comparison of the associated pulsar properties of HAWC~J2019+368 and HESS J1825-137 \citep{ATNF_catalogue,Roberts2002,Kirichenko2015}.}
\label{tab:pulsar_properties_comparison}
\end{table*}

It can be seen that both pulsars have similar $P$, $\dot{P}$, $\dot{E}$, and $\tau_c$. However, the main difference lies in the observed \gam-ray luminosities of these two systems. For a given distance of 1.8 kpc and 4 kpc for HAWC~J2019+368 and HESS J1825-137, respectively, the integrated \gam{}-ray luminosities are $\sim\!\!10^{33}$ and $\sim\!\!10^{35}$ erg s$^{-1}$, respectively. In order to match the \gam-ray luminosity of HAWC~J2019+368 to the one for HESS J1825-137, the distance to the HAWC~J2019+368 system would have to be $\sim\!\!18$ kpc. However, with the previous distance measurements this seems to be very unlikely (see Section \ref{sec:model}). In both systems, the assumed conversion efficiency of spin-down power of the pulsar to e$^{\pm}$ are below 10\% \citep{Aharonian2006a}. Assuming the radiation fields in both systems do not differ drastically, and with similar B-fields of a few $\mu$G, the energy conversion efficiency of e$^{\pm}$ to \gam-ray production should be similar for both. One possible explanation for the differing luminosities is that PSR~J2021+3651 has a larger $P_0$ (80 ms in this work) compared to PSR B1823-13 ($\sim$1 ms or $\sim$15  ms, from \cite{lowe_br_period_1825, J1825_peak_tev}). Due to the similar present day $P$ and $\dot{P}$ of the two systems and assuming $\dot{P}$ did not change rapidly in the past, PSR B1823-13 must have injected more power into the system compared to PSR~J2021+3651. 

The larger $P_0$ of PSR~J2021+3651, in turn, makes the HAWC~J2019+368 system younger ($\sim$7 kyr) compared to its characteristic age of $\sim$17 kyr. This interpretation agrees with the peak emission energies of HESS J1825-137 and HAWC~J2019+368, which are at $\sim\!\!0.1$ TeV \citep{J1825_peak_tev} and $\sim\!\!20$ TeV (Figure \ref{fig:sedJ2019}), respectively. Given the pulsars' observed similarity, the pulsar powering HAWC~J2019+368 system should be younger than the one powering HESS J1825-137 to explain the different spectral features at TeV energies: the higher peak emission energy and hard spectrum below $\sim$20 TeV. 

One can estimate the parent electron population energy for a given \gam-ray photon energy in the Thompson regime (only assuming CMB, see \cite{longair}) as:
\begin{equation}
E_e \simeq 17.2 \text{ TeV} \left( \frac{E_{\gamma {\rm -ray}}}{1 \text{TeV}}\right)^{0.5}.
\label{eqn:el_en_g-ray}
\end{equation}
For a 20 TeV photon, the parent electron population would be of energy $\sim$80 TeV. The KN effect only starts to dominate above tens of TeV of corresponding \gam{}-ray energies (assuming electrons are only scattering off of CMB photons).  Using equation \ref{eqn:cooling}, the corresponding cooling timescale for 80 TeV electrons in a 3.5 $\mu$G B-field and CMB are of the order of $\sim$9 kyr, which is similar to the estimated age of the system, $\sim$7 kyr. The electrons with energies $>80$ TeV are already cooled, making the TeV spectrum softer above \gam{}-ray energies of $\sim$20 TeV. The lower-energy electron population $<80$ TeV has not cooled yet, making the spectral index hard up to 20 TeV in \gam-ray energies. This can also be seen from Figure \ref{fig:evolved_particle_spectrum} where the break in the evolved particle spectrum can be seen at $\sim$80 TeV. 

It is also to be noted that \cite{Aliu2014} showed cone-shaped radio emission starting at the pulsar location and extending towards the centroid of TeV emission for VER~J2019+368. The positions of VER~J2019+368 (RA, Dec = $304.846\degr$, $36.779\degr$) and HAWC~J2019+368 (RA, Dec = $304.92\degr$, $36.76\degr$) are separated by $0.06\degr$, which is within the VERITAS PSF (0.1\degr{} above 1 TeV) and HAWC PSF ($0.1$\degr{}-$0.3$\degr{} for the events above 3.16 TeV depending on the bin) \citep{Aliu2014,Abeysekara2018a,Abeysekara2019}. Both observations independently support the claim that emission at TeV energies is offset from the pulsar location (RA, Dec = $305.27275\degr$, $36.85133\degr$) by approximately $0.3\degr$ \citep{Abdo2009a}. This is because the lower energy ($<80$ TeV) electrons can still emit in TeV energies via IC due to their longer cooling timescales, and within that time the pulsar had moved to its present location. On the other hand, due to the higher energies of the X-ray emitting electrons and therefore smaller cooling timescales, the X-ray emission is closer to the pulsar position \citep{Mizuno2017}. If we assume that the pulsar was born at the location of HAWC~J2019+368, and assume that the current location of the pulsar is due to only its proper motion, the transverse velocity would be $\sim$1300 km s$^{-1}$. The obtained transverse velocity is certainly at the higher end of the pulsar transverse velocity distribution \citep{transverse_velo_pulsar}, however it would not be unrealistic.

The measured extension of HAWC~J2019+368 is about $0.36\degr$ along the major axis obtained from the morphology fit using an asymmetric 2D Gaussian as described in Section \ref{sec:morphologicalandspectralfit}, which would correspond to a projected size of $\sim$11 pc. In the energy-dependent morphology study (see Section \ref{sec:energy dep morph}), energy band 3 has similar size and corresponds to the energy range between 10 and 56 TeV. The median \gam-ray  energy in this band is $\sim$23 TeV, which corresponds to electron energy of about $\sim\!\!80$ TeV (using Equation \ref{eqn:el_en_g-ray}).

Using these pieces of information, we may comment on the transport of electrons in an advection or diffusion scenario.  Recalling that the cooling timescale of mono-energetic 80 TeV electrons is $\sim\!\!9$ kyr, the corresponding advective transport speed of electrons is about $\sim$1200 km s$^{-1}$.
The Bohm diffusion coefficient is $D(E) = \eta r_g c / 3$, where $r_g = 3.3 \times 10^{15} (E/10 \rm{TeV})(B/10\mu{\rm G})^{-1} \rm {cm}$ and $c$ is the speed of light in vacuum \citep{Aharonian_book}. In the case of Bohm diffusion ($\eta = 1)$, the diffusion coefficient in a 3.5~$\mu$G B-field would be $\sim\!\!7.5\times10^{26}$ cm$^2$ s$^{-1}$.
%(page number 209). 
This value is similar to the case of Geminga and B0656+14 where the diffusion coefficient is much smaller than the ISM value derived from the boron-to-carbon ratio \citep{Abeysekara2017d}. The latter can be expressed as: $D(E) = D(10 \rm{GeV})(E/10\rm{GeV})^{\delta}$, where $D(10\rm{GeV}) \sim \!\!10^{28}$ cm$^2$ s$^{-1}$  and $\delta = 0.5$  \citep{Aharonian_book}, which would yield a value of $\sim\!\!9\times10^{29}$ cm$^2$ s$^{-1}$ for $\sim\!\!80$ TeV electrons.
%(page number 151). 

If diffusion is the mechanism dominating the transport for HAWC~J2019+368, it would imply that the magnetic field is much more turbulent than the average in the ISM, which is not surprising considering that we are in a region heavily dominated by the pulsar wind. Hence, the advection velocity or diffusion coefficient is constrained by the morphology of the TeV emission region. However, a detailed study of transport mechanisms (advection/diffusion) would require a better measurement of energy-dependent morphology. 

\section{Conclusions}
In this work, we have performed the first detailed spectral and morphological study using HAWC data of the region surrounding MGRO~J2019+37, resolving it into the sources HAWC~J2019+368 and HAWC~J2016+371. The morphology and spectrum in this work agrees well with the VERITAS measurements from \cite{Aliu2014} and \cite{Abeysekara2018a}, consisting of an elliptical Gaussian source (for HAWC~J2019+368 and VER~J2019+368) and point source (for HAWC~J2016+371 and VER~J2016+371).  The agreement between HAWC and VERITAS for the spectrum and morphology in this region is an important milestone in comparisons between IACTs and ground array detectors like HAWC. Previously, differences between measured spectra and morphology were attributed to not well understood systematic differences \citep{Abeysekara2017}. These differences may now be explainable and quantifiable (in some cases) as the difference between the extraction region used to produce the spectrum, and the morphology measured. 

The detection of HAWC~J2016+371 confirms the source VER~J2016+371. Given its very recent confirmation as a PWN via X-ray measurements, as the HAWC exposure to this source grows, it may be useful to examine the emission from this source in detail to see if its TeV emission is similar to other PWNe, or if the supernova remnant hosting the nebula may be accelerating protons to TeV energies \citep{Guest2020}. Determining the origin of the uniform background \gam-ray source in the morphology model is an important subject for future study. 

%The spectrum of HAWC~J2019+368 is now measured up to 140 TeV. 
We find that the emission from HAWC~J2019+368 together with the X-ray data is well described via emission from a young PWN system powered by a $\sim$7$\,$kyr old pulsar. This allows us to estimate the PWN properties of the system such as the magnetic field, the pulsar spin-down energy to electron conversion efficiency, and the maximum electron energy. Follow-up studies at X-ray energies may find a spatially extended emission region that might agree with the morphology measured by HAWC and VERITAS, a finding which would support the two zone model of electrons described in this work.

The energy-dependent morphology study does not show conclusive evidence in favor of shrinking size with increasing energy due to large uncertainties in the measurements. Nevertheless, it enables us to constrain the particle transport in a PWN system. This may be improved upon with more data in the future. The recent upgrade of HAWC with an outrigger array will play a crucial role by increasing HAWC's sensitivity to showers at the highest energies as it increases the instrumented area of HAWC by a factor of 4-5 \citep{latestOutrigger,outriggerReco}. 

\section{Acknowledgements}
We acknowledge the support from: the US National Science Foundation (NSF); the US Department of Energy Office of High-Energy Physics; the Laboratory Directed Research and Development (LDRD) program of Los Alamos National Laboratory; Consejo Nacional de Ciencia y Tecnolog\'ia (CONACyT), M\'exico, grants 271051, 232656, 260378, 179588, 254964, 258865, 243290, 132197, A1-S-46288, A1-S-22784, c\'atedras 873, 1563, 341, 323, Red HAWC, M\'exico; DGAPA-UNAM grants IG101320, IN111315, IN111716-3, IN111419, IA102019, IN112218; VIEP-BUAP; PIFI 2012, 2013, PROFOCIE 2014, 2015; the University of Wisconsin Alumni Research Foundation; the Institute of Geophysics, Planetary Physics, and Signatures at Los Alamos National Laboratory; Polish Science Centre grant, DEC-2017/27/B/ST9/02272; Coordinaci\'on de la Investigaci\'on Cient\'ifica de la Universidad Michoacana; Royal Society - Newton Advanced Fellowship 180385; Generalitat Valenciana, grant CIDEGENT/2018/034; Chulalongkorn University’s CUniverse (CUAASC) grant. Thanks to Scott Delay, Luciano D\'iaz and Eduardo Murrieta for technical support.

%\newpage{}
%\nocite{*}
\bibliographystyle{aasjournal}
\bibliography{HAWCJ2019.bib}

\end{document}